\documentclass[aps,prx,reprint,showpacs,amsfonts,longbibliography]{revtex4-1}

\usepackage{cmap}
\usepackage[utf8x]{inputenc}
\usepackage[T1]{fontenc}
\usepackage{textcomp}
\usepackage{amsmath}
\usepackage{latexsym}
\usepackage{float}
 \usepackage{accents}
 \usepackage{amssymb}
\usepackage{graphicx}
\usepackage{hyperref}
\usepackage{subfigure}
\usepackage{bbold}
\usepackage{xcolor}
\usepackage{url}
\usepackage{booktabs,tabularx,dcolumn}

\def\ba{\begin{eqnarray}}
\def\ea{\end{eqnarray}}
\def\be{\begin{equation}}
\def\ee{\end{equation}}
\def\bm{\begin{math}}
\def\me{\end{math}}

\def\del{\partial}

\newcommand{\dummy}

\begin{document}
\title{Activity Mediated  Globule to Coil Transition of a Flexible Polymer in Poor Solvent}
\author{ Subhajit Paul$^1$} \email[]{subhajit.paul@itp.uni-leipzig.de}
\author {Suman Majumder$^1$}\email[]{suman.majumder@itp.uni-leipzig.de}
\author{Wolfhard Janke$^1$}\email[]{wolfhard.janke@itp.uni-leipzig.de}
\affiliation{$^1$Institut  f\"{u}r Theoretische Physik, Universit\"{a}t Leipzig, IPF 231101, 04081 Leipzig, Germany}

\date{\today}

\begin{abstract}
Understanding the role of self-propulsion on the conformational properties of active filamentous objects has relevance in biology. In this context, we consider a flexible bead-spring model polymer for which along with both attractive and repulsive interactions among the non-bonded monomers, activity for each bead works along its intrinsic direction of self-propulsion. We study its kinetics in the overdamped limit, following a quench from good to poor solvent condition. We observe that with low  activities,  though the kinetic pathways remain similar, the scaling exponent for the relaxation time of globule formation becomes smaller than that for the passive case.  Interestingly, for higher activities when self-propulsion dominates over interaction energy, the polymer becomes more extended. In its steady state, the variation of the spatial extension of the polymer, measured via its gyration radius, shows two completely different scaling regimes:  The corresponding Flory exponent  changes from $1/3$ to $3/5$ similar to a transition of the polymer from a globular state to a self-avoiding walk. This can be explained by an interplay among  three energy scales present in the system, viz., the ``ballistic'', thermal, and interaction energy. 
\end{abstract}


\maketitle

\section{ Introduction}
\par
Various self-propelling objects over a diverse range can be recognized as ``active'' particles which became of significant research interest in the past two decades \cite{chate_08,roman,elgeti_15,cates1,winkler_20,shaeb_20,spreng_20}. Different theoretical models followed by various experimental approaches using synthesized colloidal particles have been involved in understanding the properties  of such systems \cite{vicsek_95, toner98,hagen_11,fily_12,farrell_12,mcand_12,mishra_12,redner_13,sten14,das_17,siebert_17,digre_18,lowen_20,paul1_21,desei12,kumar14,ravi_17,lau09,vutukuri_17,harder_14,kaiser_15,eisen_16,isele16,eisen_17,duman_18,bianco_18,paul2_20,paul3_20}. In recent years,  notable  attention has been paid to ``real'' biological systems as well \cite{schliwa03,backouche06,juelicher07,weber_15,suzuki17}.  Although the strategies for self-propulsion are not the same for ``active'' objects in different systems, emergence of various self-organized spatio-temporal patterns during their evolution, e.g., phase separation like in gas-liquid systems,  coherent collective motion with long-range order, giant density fluctuations, etc., are quite generic at all length scales \cite{roman,vicsek_95,chate_08,elgeti_15,spreng_20,mishra_12,mcand_12,farrell_12,paul1_21,winkler_20,fily_12,toner98,cates1}. For example, bacteria typically move using their filament-like cilia or flagella, whereas actin filaments of a cytoskeleton use molecular motors for their motion. On the other hand, flock of birds or herd of sheep control their motion by looking or interacting with their neighbors \cite{chate_08,vicsek_95,toner98}.
\par 
The first minimal model regarding simulation of an ``active'' system was proposed  by Vicsek et al. \cite{vicsek_95}. In the Vicsek model, the alignment activity rule provides a coherent motion of the constituents  at sufficiently high density and low external noise \cite{vicsek_95,chate_08}. Another model is a system consisting of active Brownian particles (ABP) \cite{fily_12,mishra_12,farrell_12,mcand_12,digre_18,roman,lowen_20,sten14}. For this the self-propulsion of a particle is mediated by its translational and rotational diffusion. At sufficiently high particle densities, this system shows activity induced phase separation in case of a completely repulsive interaction among them \cite{digre_18}.  Whereas variations of both these models have been vastly used in the literature to understand the behavior of particle-like systems \cite{shaeb_20,chate_08,roman,farrell_12,mcand_12,mishra_12,sten14,paul1_21,das_17,siebert_17,lowen_20,mishra_12,digre_18,redner_13}, studies related to the dynamics of filament-like entities using such active particles are rather less \cite{harder_14,kaiser_15,eisen_16,isele16,eisen_17,bianco_18,paul2_20,paul3_20,vutukuri_17,sdas_21}.
\par
Filaments and polymeric objects are integral part of many biological systems.  A  passive polymer, i.e., in absence of any activity or self-propulsion, following a quench from good to poor solvent condition  undergoes changes of its conformations from coil to a globular one \cite{doi_86,rubin_poly,byrne_95,kuznetsov_95,klushin_98,halperin_00,abrams_02,montesi_04,kikuchi_05,pham_08,guo_11,bunin_15,majumder_17,christiansen_17,majumder_20}. Kinetics of such collapse for a passive polymer has been studied extensively in the past few years, using both lattice \cite{kuznetsov_95,christiansen_17} and off-lattice models \cite{byrne_95,klushin_98,halperin_00,pham_08,bunin_15,majumder_17,majumder_20,guo_11} as well as using all-atom simulations \cite{majumder_19}. Regarding this, the phenomenological description using the ``pearl-necklace'' model by Halperin and Goldbart \cite{halperin_00} is well accepted.  Thus one asks whether such a description remains valid for active polymers as well and if not, then how the conformations and dynamical properties get affected due to  active or self-propelling forces. 
\par 
For an active polymer, its monomers can themselves be active (self-propelling) or can be activated by external forces from its environment \cite{winkler_20}.  Actin filaments or microtubules in the cell cytoskeleton are good  examples of linear filamentous objects \cite{juelicher07}. They move using the motor proteins attached to them as well as  due to external driving forces acting tangentially along their contour \cite{juelicher07,isele16,jiang_14}. In absence of hydrodynamic effects, i.e., without conservation of local momentum, the dynamics of a filament consisting of active beads is mediated by their internal noise \cite{kaiser_15,eisen_17,eisen_16}. This resembles the motion of microswimmers with strong coupling among them within a medium.  On the other hand, when a passive filament is immersed in a bath of active particles, the dynamics is governed by the force exerted on its beads due to collisions with the bath particles \cite{harder_14,eisen_17}. Hydrodynamic flows of the solvent also plays an important role on its properties \cite{jiang_14,gomez_20}. 
\par 
As mentioned, studies involving active polymers are quite new.  In experiments, a  synthesized activated polymer can be realized using a chain of colloidal or Janus particles and then making it motile by applying an electric or magnetic field \cite{yan_16,nishiguchi}. In recent years, using analytical methods and computer simulations also a few studies looked at properties of active polymers \cite{winkler_20,eisen_16,eisen_17,bianco_18,anand20,sdas_21,paul2_20,harder_14}.  Employing Brownian dynamics simulations, Bianco et al. \cite{bianco_18} have observed conformational changes of a single polymer chain consisting of active beads for which the activity on a bead works along the tangent vector of the neighboring beads. In an earlier work \cite{paul2_20}, we explored the kinetics of globule formation of a flexible polymer with increasing activity which was applied in a Vicsek-like alignment manner using underdamped Langevin dynamics. Such an activity helps in aligning the velocities of neighboring beads towards a particular direction and thus changes the pathway across the coil-globule transition.
\par 
In this paper, we conduct a comprehensive study of a coarse-grained  flexible polymer chain consisting of active Brownian beads and  look at its kinetics at  very low temperature in a poor solvent condition. Instead of using activity via any explicit alignment interaction, here, for such an active Brownian polymer (ABPo) the self-propulsion on each bead acts along its intrinsic direction. Due to the  translational and rotational diffusion of the beads and being driven by stochastic noises, the direction of propulsion changes with time.  In addition to exploring the dynamics of the polymer with increasing activity using the Langevin equation in its overdamped limit,  we also check the validity of different scaling laws known for a passive polymer in equilibrium. 

\par 
The rest of the paper is organized as follows. In Sec. II we discuss the model and methods of our simulations. Then the results are presented in Sec. III followed by the conclusions in Sec. IV.

\section{ Model and Methods}
We consider a flexible bead-spring polymer chain where the monomers are connected in a linear way.  The bonded interaction between successive monomers is modeled via the standard finitely extensible non-linear elastic (FENE) potential \cite{milchev,majumder_17,majumder_20,paul3_20}
\begin{equation}\label{fene}
V_{\rm{FENE}}(r) = - \frac{K}{2}R^2 {\rm{ln}} \bigg[ 1- \bigg(\frac{r-r_0}{R}\bigg)^2\bigg]\,,
\end{equation}
where $r_0=0.7$ is the equilibrium bond length, $K$ is the spring constant which is set to $40$ and $R=0.3$ represents the maximum extension of the bonds from its equilibrium value.
\par 
The non-bonded monomer-monomer interaction with $r$ being the spatial separation between them, is modeled via the standard Lennard-Jones (LJ) potential \cite{majumder_17,paul3_20,paul1_21}
\begin{equation}\label{lj_def}
 V_{\rm{LJ}}(r) = 4\epsilon \bigg[\bigg(\frac{\sigma}{r}\bigg)^{12}- \bigg(\frac{\sigma}{r}\bigg)^6\bigg]\,,
\end{equation}
where $\epsilon$ is the interaction strength which is set to unity, i.e., all the energies are measured in units of $\epsilon$. The bead diameter $\sigma$ is related to $r_0$ as $\sigma= r_0/2^{1/6}$. The repulsive part of the potential takes care of the excluded volume interaction. This potential has a minimum at $2^{1/6}\sigma \equiv r_0$.
\par
For advantages in numerical simulations the potential is truncated and shifted at a cut-off distance  $r_c=2.5\sigma$ such that the non-bonded pairwise interaction takes the form \cite{frenkel}
\begin{equation}\label{nb_poten}
  V_{\rm{NB}}(r)=
\begin{cases}
  V_{\rm{LJ}}(r)-V_{\rm{LJ}}(r_c) -(r-r_c)\frac{dV_{\rm{LJ}}}{dr}\Big|_{r=r_c}  r<r_c \,,\\
0 ~~~~~~~~~ \text{otherwise}\,,
   \end{cases}
\end{equation}
which is continuous and differentiable at $r=r_c$, and has the same qualitative behavior as $V_{\rm{LJ}}$.
\par
The dynamics of the ABPo chain is studied using overdamped Langevin equations. The activity for each bead works along its direction of self-propulsion which changes with time. For each active bead we work with the following equations for the translational and rotational motion \cite{kaiser_15,digre_18,lowen_20,sten14},
\begin{equation}\label{trans}
\del_t{\vec{r}}_i = \beta D_{\rm{tr}}[-\vec{\nabla} U_i+f_p \hat{n}_i]+\sqrt{2D_{\rm{tr}}}\,\vec{\Lambda}_i^{\rm{tr}},\\
\end{equation}
and
\begin{equation}\label{rot}
\del_t{\hat{n}}_i = \sqrt{2D_{\rm{rot}}}(\hat{n}_i\times \vec{\Lambda}_i^{\rm{rot}}).
\end{equation}  
Here $\vec{r}_i$ represents the position of the $i$-th particle, $U_i$ is the passive interaction consisting of both $V_{\rm{FENE}}$ (between bonded monomers) and $V_{\rm{NB}}$ (among the non-bonded monomers), $\beta=1/k_BT$ is the inverse temperature, and $f_p$ (same for all beads and constant over time) denotes the strength of the self-propulsion force acting along the unit vector $\hat{n}_i$. `$\times$' represents the cross-product between two vectors.
 $\vec{\Lambda}_i^{\rm{tr}}$ and $\vec{\Lambda}_i^{\rm{rot}}$ are the random noises with zero-mean and unit-variance and are Delta-correlated over space and time given by
\begin{equation}\label{noise_correl}
\langle\Lambda_{i\mu}(t)\Lambda_{j\nu}(t')\rangle = \delta_{ij}\delta_{\mu \nu} \delta(t,t')\,,
\end{equation}
where $i,j$ are the particle indices and $\mu, \nu$ represent the Cartesian components.  In Eqs.~(\ref{trans}) and (\ref{rot}) $D_{\rm{tr}}$ and $D_{\rm{rot}}$ are, respectively, the translational and rotational diffusion constants of the beads.  Their relative importance is defined as \cite{elgeti_15,eisen_16,sten14,lowen_20}
\begin{equation}\label{delta}
	\Delta = \frac{D_{\rm{tr}}}{D_{\rm{rot}}\sigma^2}\,.
\end{equation}
In our simulation we have fixed $\Delta = 1/3$ \cite{eisen_16}. $D_{\rm{tr}}$ is related to $\beta$ and the drag coefficient $\gamma$ as $D_{\rm{tr}}=1/ \gamma \beta$, where we have chosen $\gamma=1$. $\del_t$  represents the first-order time derivative.  Time is measured in units of $\sigma^2\gamma/\epsilon$ ( $\propto 1/D_{\rm{rot}}=\sigma^2 \gamma \beta$). For the simulations, we have set the integration time step to $ 10^{-5}$ in units of this timescale.  
\par 
Following the literature, a dimensionless quantity, the P\'eclet number is defined as the ratio between the strength of the active force $f_p$ and the thermal force $k_BT/\sigma$ as \cite{elgeti_15,eisen_16,sten14,lowen_20}
\begin{equation}\label{pe_def}
 {Pe}=f_p\sigma/k_BT.
\end{equation}
By choosing $T=0.1$ we keep the thermal noise small.
For the passive case, $Pe=0$, this $T$ is well below the $\Theta$-transition temperature $T_{\theta}=2.646(4)$ \cite{majumder_17}. In the rest of the paper the results for different strength of the activity are presented in terms of $Pe$.

\par
For both passive and active cases, the initial configurations are prepared at a high temperature where the polymer is in an extended coil state. 
Then the time evolution of this polymer is studied by tuning the activity parameter $f_p$.  We considered chains with the number of beads $N$ varying over a wide range between $32 \le N \le 380$ and P\'eclet numbers between $0 \le Pe < 70$. $\langle \dots \rangle$ represents averaging over $100$ independent realizations (initial conditions and thermal noise).

\section{ Results}


We organize the presentation of our results into two subsections. In the first, we investigate the nonequilibrium kinetics of the polymer during its evolution towards the corresponding steady state. The properties of the steady-state conformations are then discussed in the second subsection.

\subsection{Nonequilibrium kinetics}


\par
\begin{figure*}[t!]
	\centering
	\includegraphics*[width=14.2cm, height=15.1 cm]{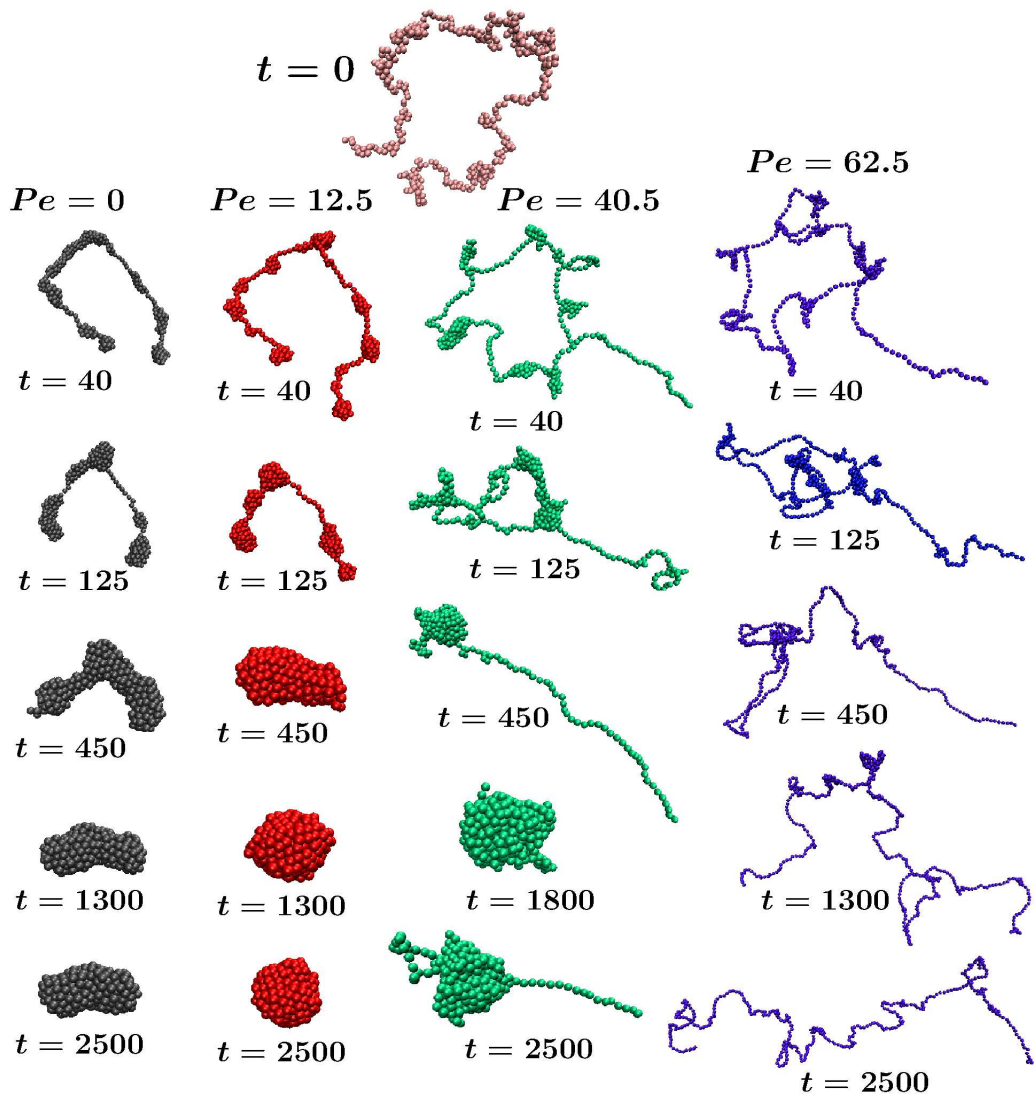}
	\caption{\label{snaps} Conformations at different times $t$ (in units of $\sigma^2\gamma/\epsilon$) during the evolution of a polymer with $N=380$ for the passive ($Pe=0$) as well as for a few active cases. Snapshot corresponding to $t=0$ represents the starting conformation.}
\end{figure*}

 
In Fig.~\ref{snaps} we show the typical nonequilibrium conformations for passive as well as a few active cases of the polymer. For all the cases we started at $t=0$ with an extended conformation of the polymer.  As observed in a few earlier studies, the globule formation of the passive polymer occurs via  ``pearl-necklace'' like arrangements of the monomers along the chain \cite{klushin_98,halperin_00,majumder_17,christiansen_17,majumder_20}.  According to this description, in the first stage, after the quench,  a few small clusters appear uniformly  along the chain. During the second stage, which is known as the coarsening stage, the clusters grow in size when the monomers from their connecting bridges move diffusively and join them. Then as the lengths of the bridges decrease, these clusters meet each other and eventually merge to form a larger one. This process continues until a single cluster or globule forms. In the final stage, the beads within this  globule rearrange themselves and form a more compact structure to minimize its surface energy. Here, after the formation of the globule, only the thermal fluctuations drive the rearrangement process of the beads in order to make the globule more spherical.

\begin{figure}[b!]
	\centering
	\includegraphics*[width=8.5cm, height=7.16cm]{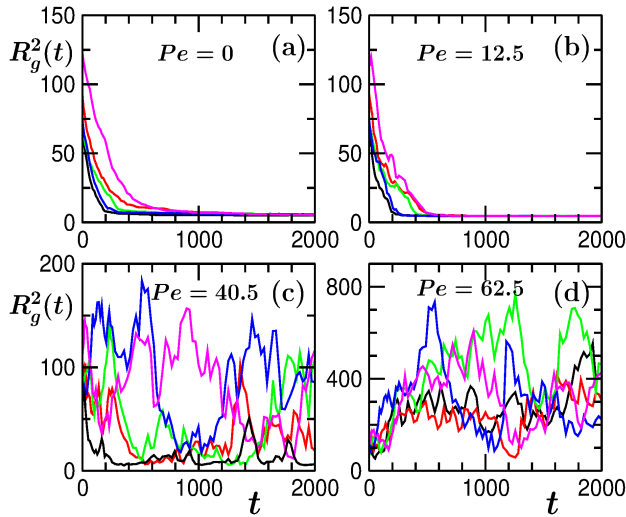}
	\caption{\label{rg_t_runs} Plots of $R_g^2$ versus $t$ for a few typical individual runs with $N=380$ for different values of $Pe$ during time evolution of the polymer.}
\end{figure}

\begin{figure*}[t!]
	\hskip -0.7cm
	\includegraphics*[width=0.348\textwidth]{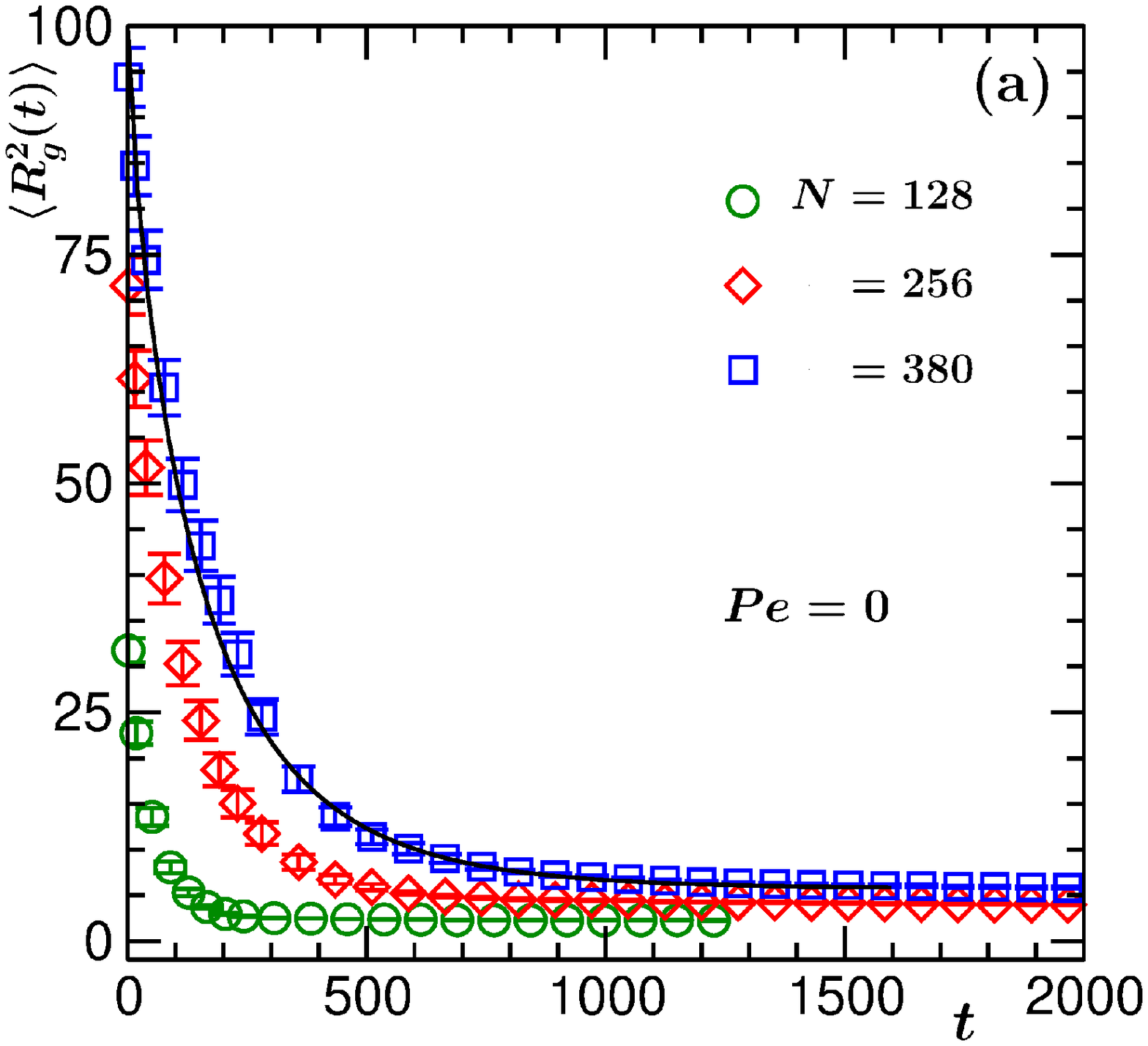}
	\hskip -0.19cm
	\includegraphics*[width=0.345\textwidth]{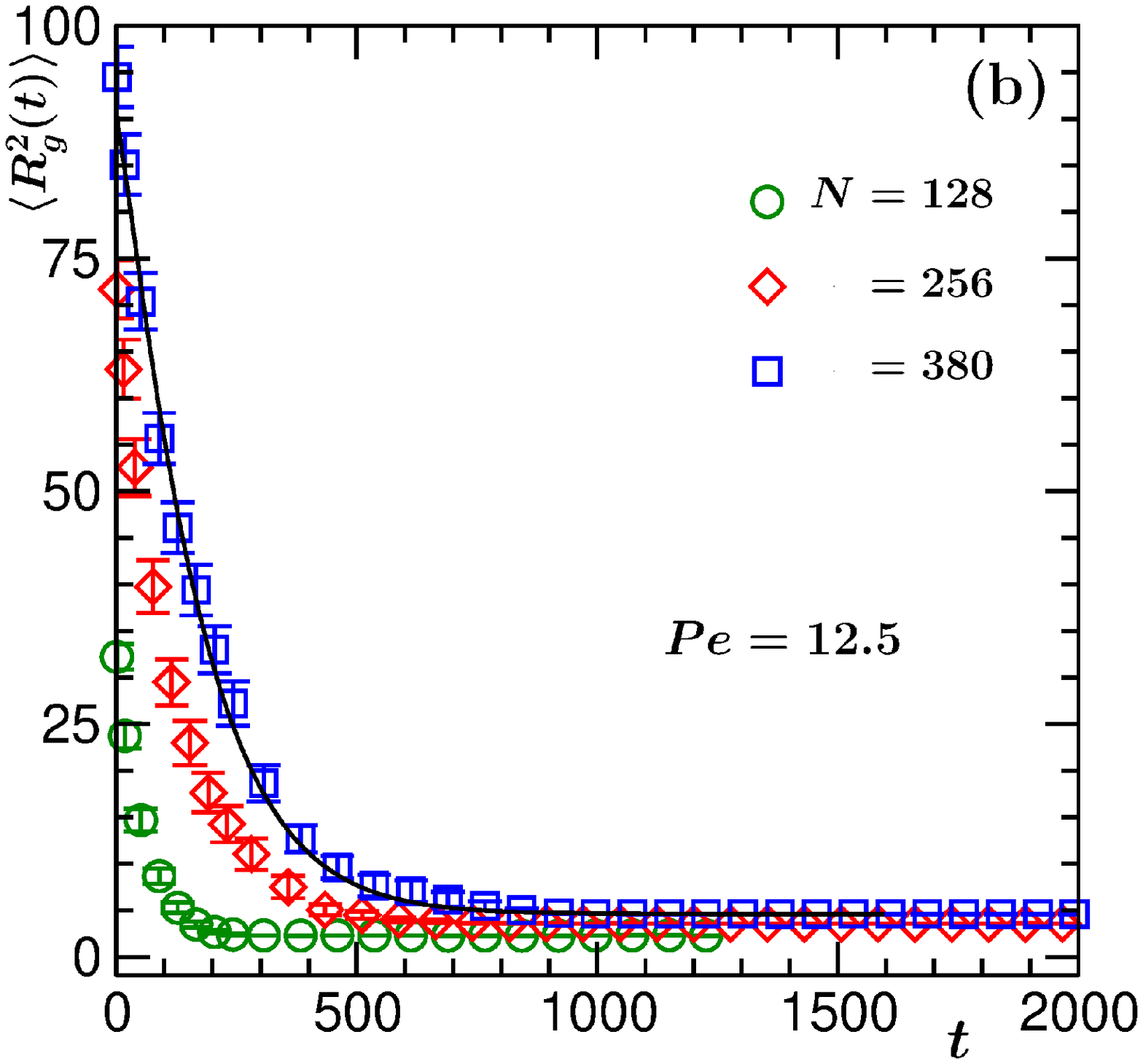}
	\hskip -0.18cm
	\includegraphics*[width=0.355\textwidth]{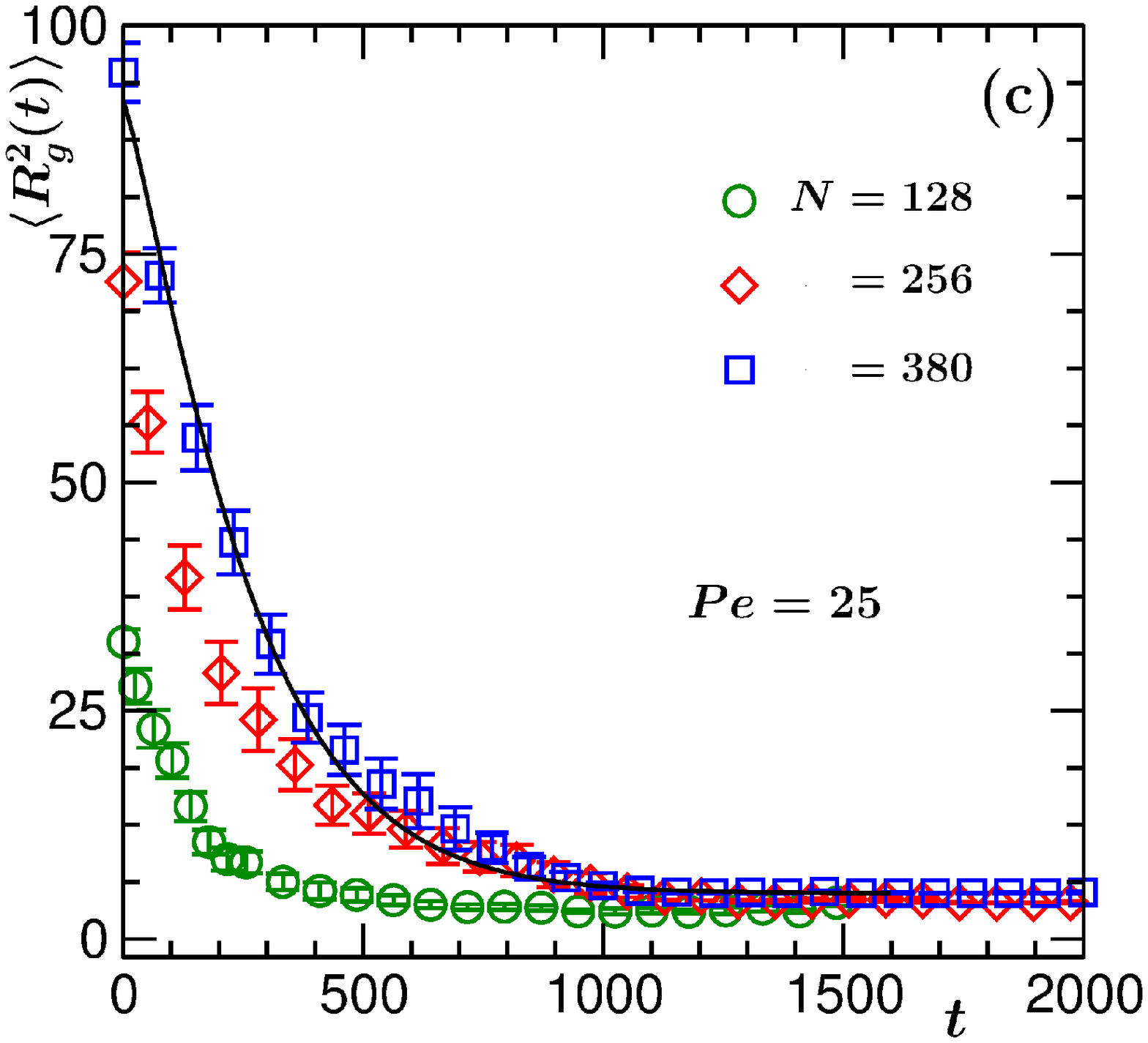}
	\caption{\label{rgsq_t} Plots of squared radius of gyration $\langle R_g^2(t)\rangle$ versus $t$ for (a) the passive case ($Pe=0$) and active chains with (b) $Pe=12.5$ and (c) $Pe=25$, for three different chain lengths.}
\end{figure*}
\par 
Coming to active polymers, for low activity strength with $Pe= 12.5$, as evident from Fig.~\ref{snaps}, the conformations are somewhat similar to those for the passive case. One point to notice is that the final conformation looks somewhat more spherical. Now for $Pe=40.5$  the conformations of the ABPo look quite different than in the earlier cases. The intermediate ``pearl-necklace''-like structures do not appear. Instead the polymer makes a ``loop-like'' structure. Then the end of the polymer where the ``loop'' has formed, coarsens faster than the other end resulting in a ``head-tail''-like conformation. The ``head'' gradually increases in size by the addition of monomers from the ``tail'' part and eventually all the monomers become  part of a single cluster. Proceeding further in time we observe that the polymer again  slowly starts stretching from both the ends. For a much higher activity, with $Pe= 62.5$, the conformations at early times look quite similar to the $Pe=40.5$ case. Then with increasing time we see that the polymer stretches more and more instead of collapsing. In this case, the final steady-state conformation of  the polymer becomes a completely extended one.

For the quantification of kinetics across the coil-globule transition,  we measure the squared radius of gyration defined as
\begin{equation}\label{rgsq_def}
R_g^2 = \frac{1}{N}\sum_{i=1}^{N} (\vec{r}_{\rm{cm}} - \vec{r}_i)^2,
\end{equation}
where $\vec{r}_{\rm{cm}}$ is the center-of-mass of the polymer given by
\begin{equation}\label{com_def}
\vec{r}_{\rm{cm}}=\frac{1}{N}\sum_{i=1}^{N} \vec{r}_i.
\end{equation}
From the average value of $R_g^2$ it is not possible to understand the fluctuations related to the conformations in each run. As an illustration we show in Figs.~\ref{rg_t_runs}(a)-(d) the variation of $R_g^2$ versus $t$ for a few typical runs with $N=380$ at the $Pe$ values used in Fig.~\ref{snaps}. For $Pe=0$ and $12.5$, $R_g^2$ decays monotonically and the individual runs mainly exhibit an up-down shift. Smaller values of $R_g^2$ close to $5$ for the runs indicate the formation of a globule. For $Pe=40.5$ one observes an oscillatory behavior  for any individual run. After reaching a compact globule the structure again starts to extend due to activity of the beads, as seen from Fig.~\ref{snaps}. Such oscillatory behavior can not be observed in averaged data as the corresponding  times for collapse and re-expansion are not the same for different runs. Now for $Pe=62.5$ the polymer always extends from its starting point. This can be understood by looking at the values of $R_g^2$ on the $y$-axis.
\par 

Next in Fig.~\ref{rgsq_t}(a) we plot the average $\langle R_g^2 \rangle$ versus $t$ in the passive case $Pe=0$, for three different values of the chain length.  We see that the decay of $R_g^2$ becomes slower with increasing $N$  as observed earlier \cite{majumder_17,majumder_20,paul2_20}. To check for the active cases, in Figs.~\ref{rgsq_t}(b) and (c) we plot $\langle R_g^2 \rangle$ versus $t$ for $Pe=12.5$ and $25$, respectively. For these also, we see a similar trend for the decay of $ \langle R_g^2 \rangle$ for different $N$ as observed for the passive case. 

\begin{table}[b!]
	\caption{Parameters $b_0$, $b_1$, $\tau_c$, and $\beta$  for different values of $Pe$, obtained by fitting the time dependence of the squared radius of gyration  $\langle R_g^2 \rangle$ with the ansatz (\ref{rgsq_fit}). All data are for $N=380$.}\label{tab1}
	\centering
	\begin{tabular}{|c|c|c|c|c|}
		\hline
		~$Pe$~~&~~$b_0$~~&~~$b_1$~~&~~$\tau_c$~~&~~$\beta$~~\\
		\hline
		~~0~~&~~5.77(9)~~&~~94(3)~~&~~148(10)~~&~~0.78(3)~~\\
		~~6.25~~&~~4.71(2)~~&~~92(2)~~&~~119(9)~~&~~0.79(3)~~\\
		~~12.5~~&~~4.59(2)~~&~~86(3)~~&~~177(10)~~&~1.15(3)\\
		~~18.75~~&~~4.75(2)~~&~~85(2)~~&~~207(12)~~&~1.28(5)\\
		~~25~~&~~5.03(2)~~&~~87(3)~~&~~264(19)~~&~~1.23(9)~~\\
		~~31.25~~&~~7.1(7)~~&~~90(3)~~&~~386(34)~~&~~1.09(9)~~\\
		\hline
	\end{tabular}
\end{table}

\par 
To have a comparison regarding the relaxation of the ABPo towards its final steady state for different $Pe$, we fit our data with the form \cite{majumder_17,paul2_20}
\begin{equation}\label{rgsq_fit}
	\langle R_g^2(t) \rangle=b_0+b_1 \exp[-(t/\tau_c)^{\beta}]\,,
\end{equation}
where $b_0$ is the value of $\langle R_g^2 \rangle$ at $t \to \infty$, i.e., the size of the polymer in its collapsed conformation, $\tau_c$ measures the required relaxation time to reach the collapsed state, and $b_1$ and $\beta$ are fitting parameters where $b_1$ is related to $R_g^2$ at $t=0$. For $\beta=1$, Eq.\ (\ref{rgsq_fit}) corresponds to a purely exponential decay. The solid lines in Figs.~\ref{rgsq_t}(a)-(c) for $N=380$ show the resulting fits using the ansatz (\ref{rgsq_fit}). In Table~\ref{tab1} we quote values of the fitting parameters for a few values of $Pe$. The errors for the fitting parameters are calculated using the  Jackknife resampling method \cite{efron}. In all cases, the fitted $\beta$ values differ from $1$. For very low activity, i.e., $Pe=6.25$, we see that $\tau_c$ is smaller than that for the passive case, indicating a much faster relaxation to the globule. Now with increasing $Pe$ the values of $\tau_c$ become larger showing a nonmonotonic behavior of the relaxation time with the variation of $Pe$. Larger error bars for $\tau_c$ with $Pe \ge 25$  are due to the presence of larger fluctuations in $R_g^2$.  Interestingly, with increasing $Pe$ one also observes a nonmonotonicity in the values of $b_0$ which is related to the size of the globule in its steady state. Larger values of $b_0$ indicate  that the globules are not completely spherical.

\par
\begin{figure}[t!]
	\centering
	\includegraphics*[width=0.48\textwidth]{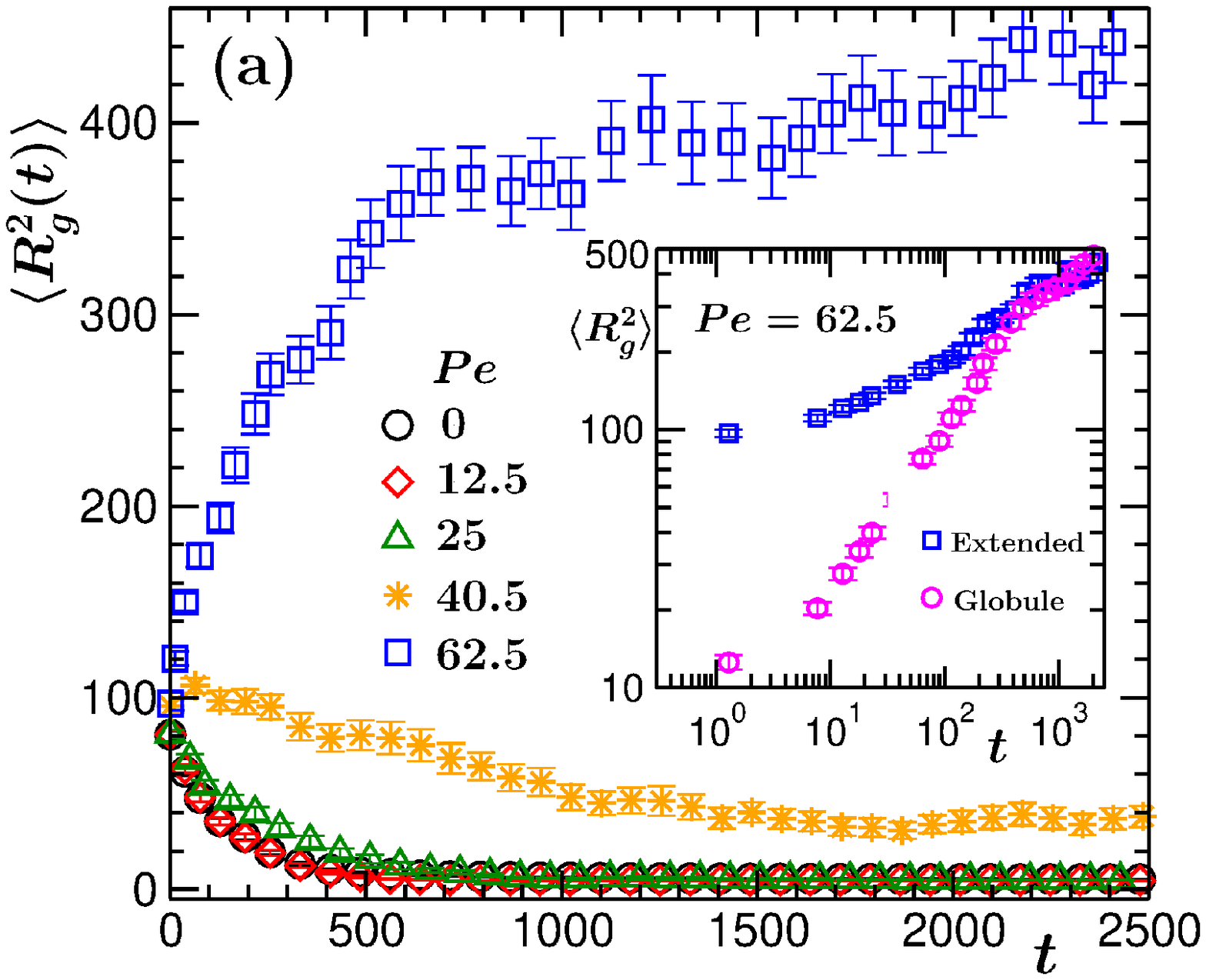}
	\vskip 0.4cm
	~~~~\includegraphics*[width=0.44\textwidth]{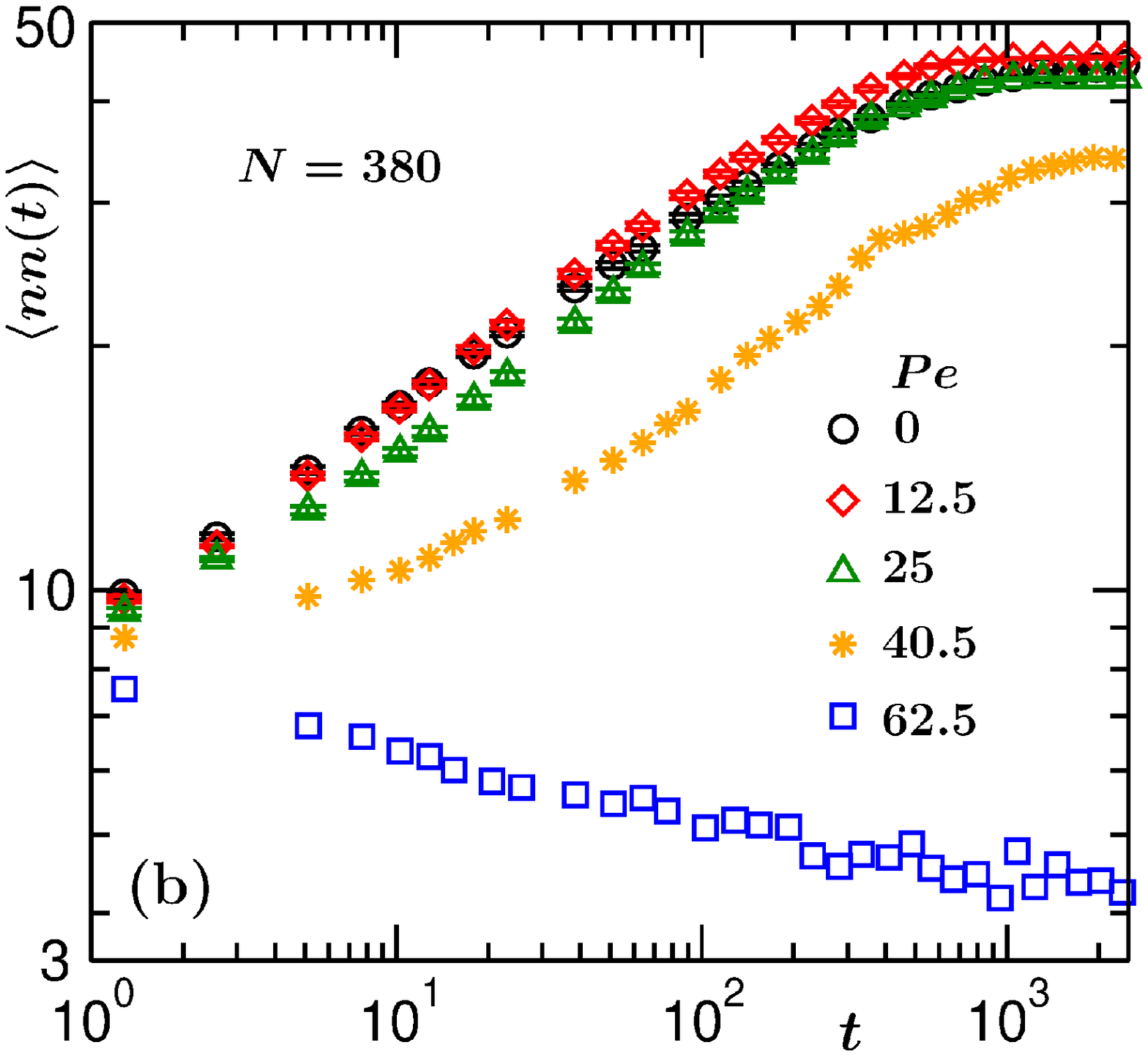}
	\caption{\label{rgsq_pe} (a) Plot of $\langle R_g^2(t)\rangle$ versus $t$ for the passive and a few active cases. Inset shows data sets only with $Pe=62.5$, for which different colors mark the cases with starting conformations being extended or a globule one.  (b) Log-log plot of the average coordination number $\langle nn \rangle$ of a monomer versus $t$ for the same $Pe$ values as in (a). All data sets are for $N=380$.} 
\end{figure}

 
Now we consider the cases with higher activities, i.e.,  $Pe=40.5$ and $62.5$.  As a qualitative comparison among different activities, in Fig.~\ref{rgsq_pe}(a) we plot $\langle R_g^2 \rangle$ versus time for different $Pe$ values for our longest considered chain, i.e., $N=380$. For $Pe=40.5$ the  decay of $\langle R_g^2 \rangle$ is much slower. For much higher activity with $Pe=62.5$, unlike the other cases, $\langle R_g^2 \rangle$ increases from its starting point, reflecting that the steady-state conformation is an extended coil (cf.~Fig.~\ref{snaps}). Now, we ask whether a stretched conformation of the polymer is a generic feature for a high enough  activity. To check for that, in the inset, we plot $\langle R_g^2 \rangle$ versus $t$ for $Pe=62.5$ with two different starting conditions. Along with the data for the extended case similar to the main frame, we also present data for which the starting conformations of the ABPo are globules. Data are presented on a log-log scale to emphasize the differences of the values of $\langle R_g^2 \rangle$ at small $t$. We see that, even after starting from entirely different initial conformations, $\langle R_g^2 \rangle$ for both of them converges towards a similar value at large $t$. It confirms that  for high enough $Pe$,  when activity dominates over the inter-monomer interactions as well as the thermal fluctuations, the steady-state  conformation of the polymer becomes an extended coil.

\par 
As another measure of such conformational change due to activity, in Fig.~\ref{rgsq_pe}(b), we plot the average coordination number $\langle nn \rangle$ of a monomer versus time $t$ for different values of $Pe$ on a log-log scale. $nn$ is calculated by counting the number of beads within the cut-off radius $r_c=2.5\sigma$, around a monomer. For passive as well as for lower activities we see that $\langle nn \rangle$ increases with time and saturates for $t \ge 500$. This indicates the formation of a globule. For $Pe=40.5$, $\langle nn \rangle$ also increases but saturates at a smaller value.  With higher activity $Pe=62.5$, on the other hand, $\langle nn \rangle$ decreases monotonically  which suggests that the polymer becomes extended. As already mentioned, the activity works on each bead along its direction of self-propulsion and thus for very high activities the beads try to move far apart from each other and essentially the polymer becomes elongated.

\begin{figure}[t!]
	\centering
	\includegraphics*[width=8.2cm, height=7.0cm]{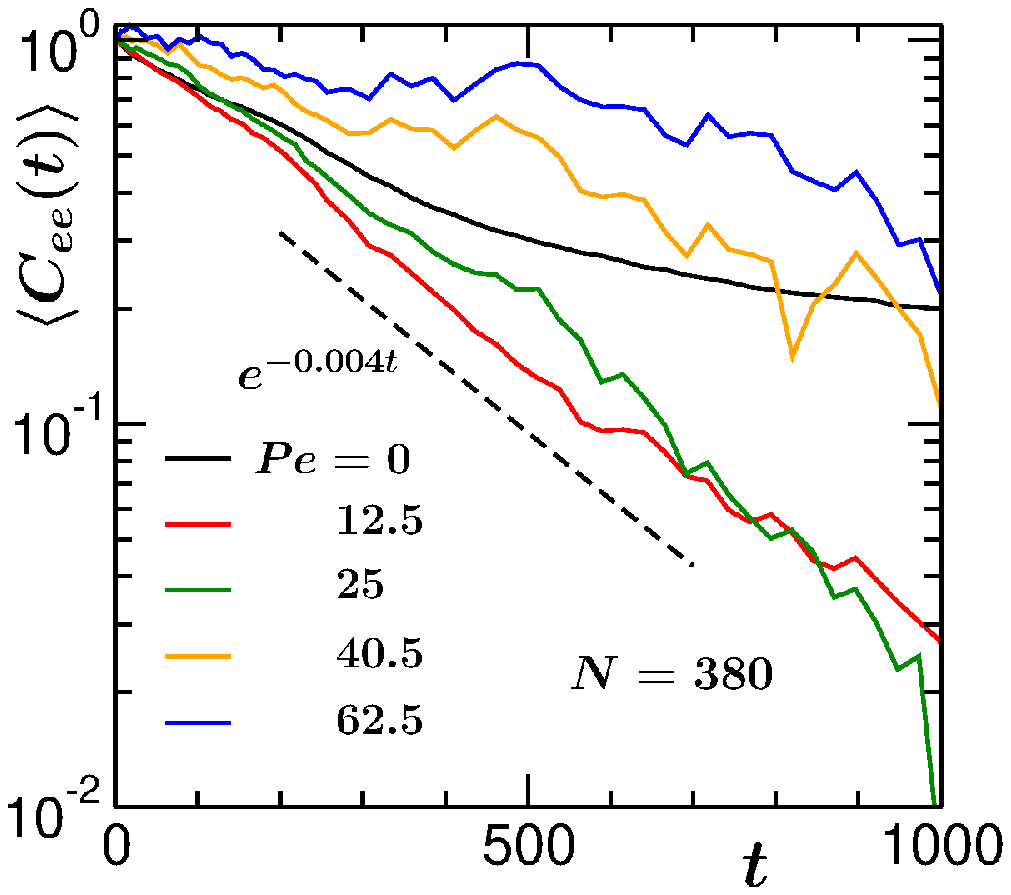}
	\caption{\label{end_vec_correl} Semi-log plot of the normalized end-to-end vector correlation $\langle C_{ee}(t) \rangle$ versus $t$, on a semi-log scale, for different values of $Pe$. All data are for $N=380$. The dashed line shows an exponential function $\exp(-0.004t)$ as a guide to the eye.}
\end{figure}
\par 
To see the effect of increasing activity on the macroscopic conformational changes with time, we calculated the end-to-end vector correlations and compared them with those for the passive case. This two-time correlation is defined as \cite{eisen_17}
\begin{equation}
C_{ee}(t)= \vec{r}_{ee}(t) \cdot \vec{r}_{ee}(0) / \vec{r}^{~2}_{ee}(0)\,,
\end{equation}
where $\vec{r}_{ee}$ stands for the end-to-end vector of the polymer given by
\begin{equation}
\vec{r}_{ee}=\vec{r}_1-\vec{r}_N\,.
\end{equation}
In Fig.~\ref{end_vec_correl} we plot this normalized end-to-end vector autocorrelation $\langle C_{ee}(t) \rangle$ versus $t$ on a semi-log scale for the passive as well as for a few active cases. For all of them, starting from a coil conformation, we measure how rapidly the conformations change with time for different values of $Pe$. For lower activities ($Pe=12.5$ and $25$) $\langle C_{ee}(t)\rangle$ follows quite a similar trend as for the passive case until $t \approx 200$. This could be expected from Fig.~\ref{snaps} as the conformations for $Pe=0$ and $12.5$ look visually quite similar.  After that initial regime $\langle C_{ee}(t)\rangle$ for $Pe=12.5$ deviates from the passive case and the decay becomes faster. In this regime, the decay is exponential-like  and  for $200 \le t \le 700$ data look consistent with $\exp(-0.004t)$. With lower activities, though the final conformations are globules and the decay of $\langle R_g^2 \rangle$ follows a  similar trend as that for the passive case, the beads can rearrange themselves within the cluster more rapidly due to their self-propulsion. This certainly helps in modifying the structure of the ABPo and leads to a faster decay of  $\langle C_{ee}\rangle$.  
Now for $Pe=40.5$  the decay of $\langle C_{ee}(t)\rangle$ is always slower compared to the passive and the lower activity  cases. This is due to the intermediate  conformational changes as observed in Fig.~\ref{snaps}.  With much higher activity, i.e., for $Pe=62.5$, the decay becomes even slower, as the polymer always remains in the extended state.  For the passive case and also for the higher activities it seems like there are continuous changes in the slopes of the decay.

\begin{figure}[t!]
	\centering
	\includegraphics*[width=0.45\textwidth]{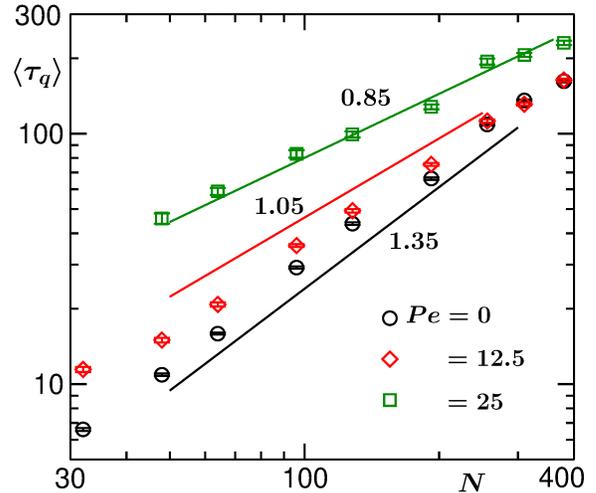}
	\caption{\label{tauc_N} Log-log plot of $\langle \tau_q \rangle$, estimated from Eq.~(\ref{relax_time}), versus chain length $N$ for the passive ($Pe=0$) and the active cases with $Pe = 12.5$ and $25$. Solid lines represent the power law \eqref{tau_scl} for which the corresponding exponents are mentioned next to them.}
\end{figure}

\par 
From the previous plots we now have a good comparative picture of the nonequilibrium pathways towards the steady-state conformations of the ABPo for low and high values of $Pe$. Further, we investigate the scaling of the nonequilibrium relaxation time for the collapse. For this, we consider the lower values of $Pe$ for which the final state of the ABPo is a globule.  
Following earlier works \cite{paul2_20,majumder_17}, such a characteristic time $\tau_q$ can be estimated from the decay of $R_g^2$ as
\begin{equation}\label{relax_time}
R_g^2 (t=\tau_q) = q\Delta R_g^2 + R_g^2 (t \to \infty)\,,
\end{equation}
where  $R_g^2( t\to \infty)$ denotes the value of the gyration radius in the globular state and $\Delta R_g^2 = R_g^2(0)-R_g^2(t \to \infty)$ measures the total decay. $\tau_q$ is the time at which $R_g^2$ decays to a fraction $q$  of its total decay. This definition is valid for the cases for which $R_g^2$ decays from its starting value.  In our analysis we have taken $q=1/e$. This choice of $q$ is motivated by the exponential-like behavior of the fitting ansatz in Eq.~(\ref{rgsq_fit}) for the data shown in Fig.~\ref{rgsq_t}. A more detailed description regarding this can be found in Refs.~\cite{paul2_20,majumder_20}. In Fig.~\ref{tauc_N} we  plot $ \langle \tau_q \rangle$ versus $N$ for the passive  as well as for the active cases with $Pe = 12.5$ and $25$. For all the cases, data show power-law behaviors as
\begin{equation}\label{tau_scl}
	\langle \tau_q \rangle \sim N^z\,,
\end{equation}
with $z$ as a dynamical exponent \cite{klushin_98,majumder_17,majumder_20}. For the passive polymer case we see $z \simeq 1.35$. Our estimated value for $z$ matches quite well with a few earlier results using Brownian dynamics simulations in absence of hydrodynamics \cite{klushin_98,pham_08,abrams_02}. However, the dynamics  appears to be faster compared to Monte Carlo simulations \cite{majumder_17,majumder_20}. 
Now coming to the active cases, with lower activities the exponent $z$ appears to be smaller compared to the passive case. Our estimated values for the exponent $z$ are $\simeq 1.05$ and $\simeq 0.85$, for $Pe=12.5$ and $25$, respectively. This indicates that for lower activities the dynamics of collapse becomes faster than that for the passive case. This scenario is in contrast to the collapse dynamics observed for a polymer with the Vicsek-like alignment activity among the monomers \cite{paul2_20}.
\par 

\subsection{Steady-state properties}
\par
\begin{figure}[t!]
	\centering
	\includegraphics*[width=0.44\textwidth]{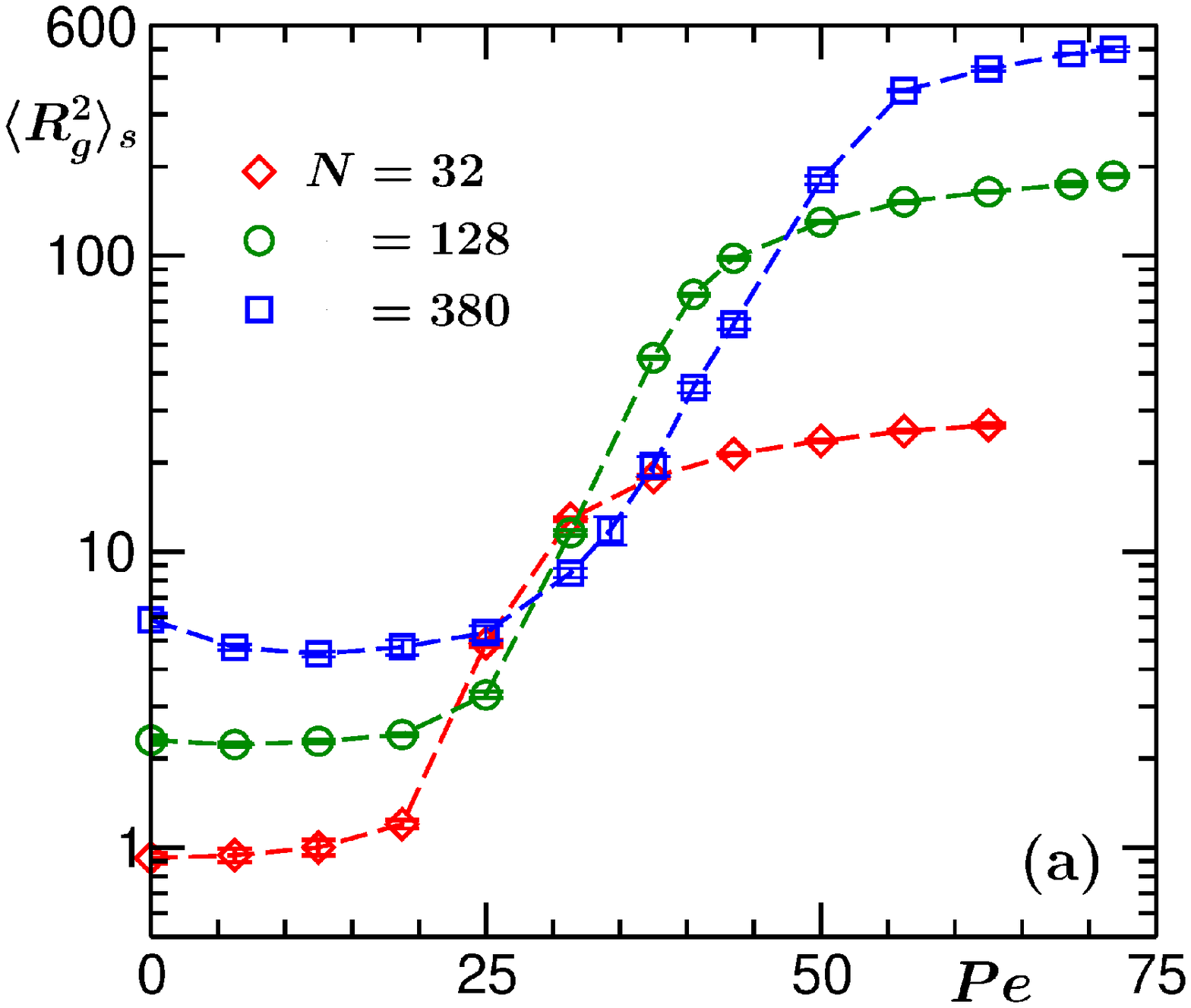}
	\vskip 0.4cm
	\includegraphics*[width=0.44\textwidth]{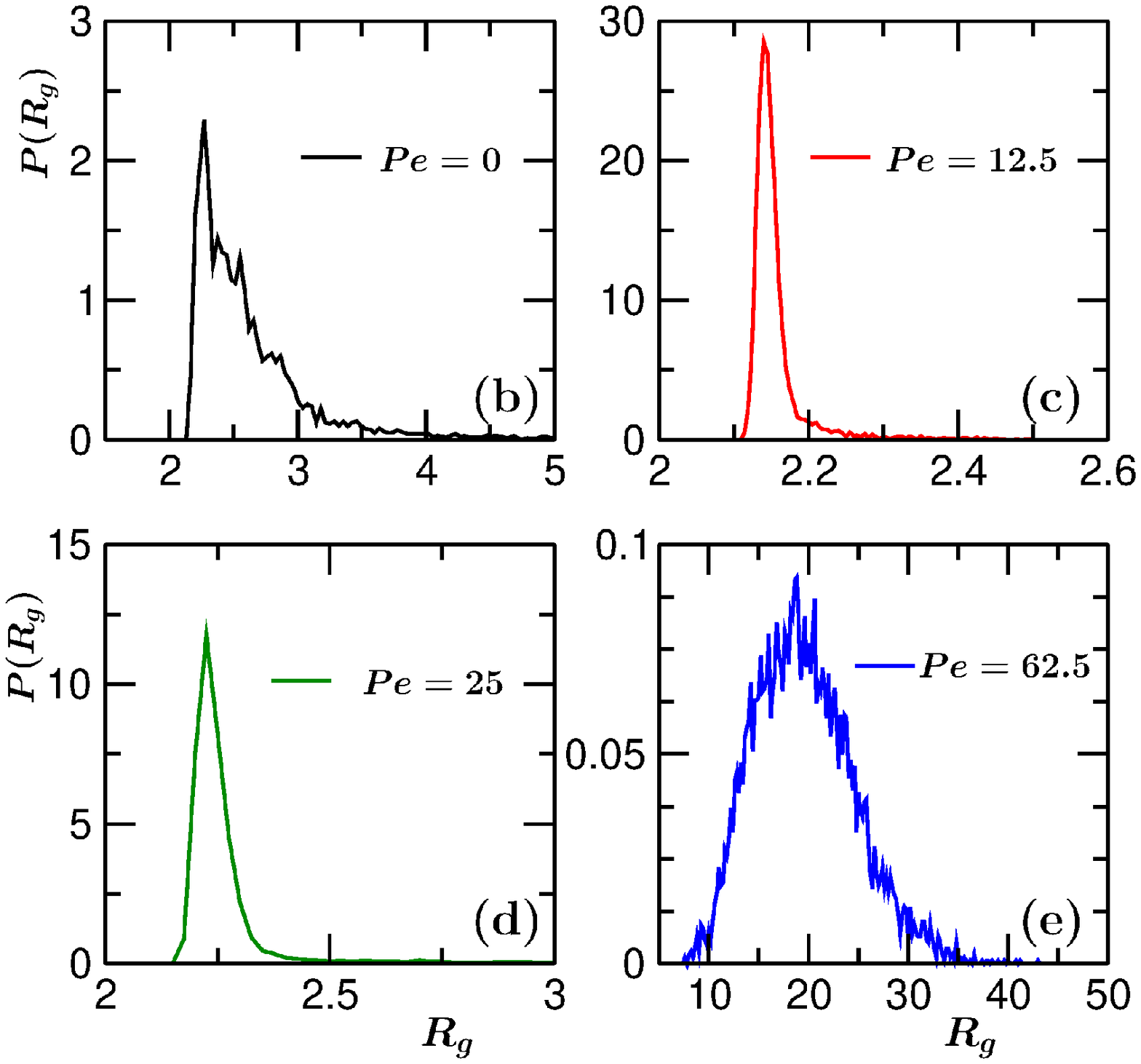}
	\caption{\label{length_ss_pe} (a) Plot shows the variation of the size of the polymer, $\langle R_g^2\rangle_s$ for its \textit{pseudo-equilibrium} steady-state  conformations, versus $Pe$. (b)-(e) Plots of the normalized distributions of radius of gyration $P(R_g)$ versus $R_g$ for the steady-state conformations at different values of $Pe$ for $N=380$.}
\end{figure}

 
Next we investigate the properties of the steady state which is reached once $\langle nn \rangle $ has saturated. In particular we study
how the size of the polymer and its related scaling changes with activity. In Fig.~\ref{length_ss_pe}(a) we plot $\langle R_g^{2} \rangle_s$ (subscript $s$ indicates steady-state averages) versus $Pe$ over a wide range, for three different chain lengths, i.e., $N=32$, $128$ and $380$, on a semi-log scale. The behavior of $\langle R_g^{2} \rangle_s$ is indeed representative of a crossover from a globule to extended conformations with increasing $Pe$. For $N=380$, $ \langle R_g^{2} \rangle_s$ first slightly decreases and then increases as a function of $Pe$. A similar nonmonotonic feature can also be identified from  the values of the fitting parameter $b_0$ in Table~\ref{tab1}.  
From this plot one can identify  the corresponding crossover points of $Pe$ which mark the onsets of the change of conformations from globule to the extended state for any $N$.  This crossover point $Pe$ increases with $N$. Such a dependence has similarity with the identification of the $\Theta$-transition temperature with the variation of $N$ in the case of a passive polymer \cite{majumder_17}.
\par 
For conformational changes of a polymer  it is  also an usual practice to look at the distribution $P(R_g)$ of gyration radius for the steady states. In Figs.~\ref{length_ss_pe}(b)-(e) we plot the normalized $P(R_g)$ versus $R_g$ at different values of $Pe$. There one observes a nonmonotonic behavior regarding the width of the distribution with increasing $Pe$: It decreases for lower activities compared to $Pe=0$ and then increases for  higher $Pe$. For passive as well as for lower activities the peaks are at $R_g \approx 2.1-2.2$, whereas for $Pe=62.5$ it shifts to the right, i.e., towards a much larger value of $R_g \approx 20$. This behavior of $P(R_g)$ confirms a  transition of the polymer from globule to extended state with increasing $Pe$.
Similar behavior of $P(R_g)$ is expected for a passive polymer as well for temperatures below and above its $\Theta$-transition temperature.

\par
\begin{figure}[t!]
	\centering
	\includegraphics*[width=0.46\textwidth]{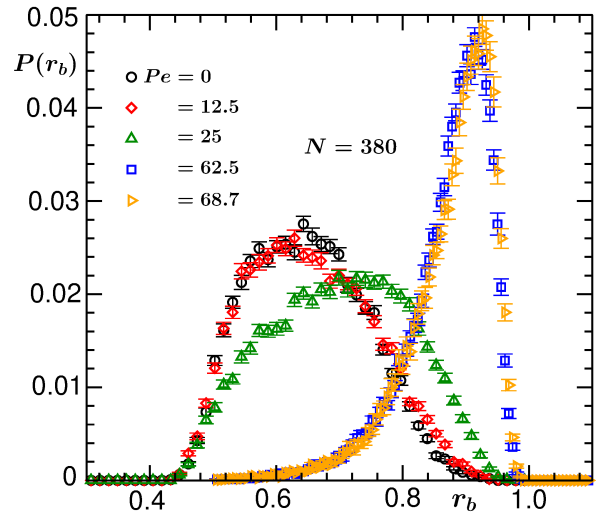}
	\caption{\label{rb_pe} Plots of the normalized distributions of bond lengths $P(r_b)$ versus $r_b$ for different values of $Pe$ for the steady-state conformations.}
\end{figure}

Until now we have seen the effect of activity in the macroscopic changes of the polymer. To see the effect in the microscopic details we measure the individual bond lengths $r_b$ between any two successive beads, defined as,
\begin{equation}
r_b =|\vec{r}_{i+1}-\vec{r}_i|\,,
\end{equation}
where $\vec{r}_i$ represents the position of the $i$-th bead, and looked at their distributions.  In Fig.~\ref{rb_pe} we plot the normalized  distributions $P(r_b)$ for different values of $Pe$ with $N=380$  for the steady-state conformations. We see that the distributions for the lower activities are  quite similar to that for the passive case. For much higher activities, i.e., with $Pe=62.5$ and $68.7$ the distributions look very different. They become more asymmetric around their mean.  Also the heights of the distributions increase and their widths decrease indicating lesser fluctuations in the variation of $r_b$. The shift of the peak positions of $P(r_b)$ to the right indicates the increase of the average bond length for the conformations.  For lower values of $Pe$ the peak occurs around $r_b \approx r_0=0.7$ whereas for higher $Pe$ the corresponding values of $r_b$ appear to be $\approx 0.92$, slightly smaller than the upper limiting value $r_0+R=1.0$ of the FENE bonds. 

\begin{figure}[t!]
	\centering
	\includegraphics*[width=0.46\textwidth]{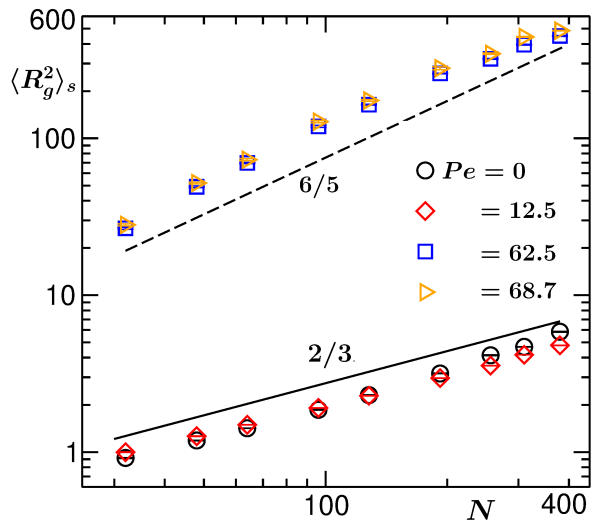}
	\caption{\label{rgsq_ss_N} Plots of the steady-state values of the squared gyration radius $\langle R_g^2\rangle_s$ of the polymer versus the chain length $N$ for the passive and a few active cases. The continuous and dashed black lines correspond to power-laws for which the exponents are mentioned next to them.}
\end{figure} 
\par 
 As observed, the steady state changes from a globular to an extended one with increasing $Pe$.  We already got an idea regarding this from Fig.~\ref{rgsq_pe}(a) which shows that $Pe=62.5$ is high enough for our considered chain lengths here (i.e., up to a maximum of $N=380$) to overcome the  attractive forces and to make the conformations more extended. We want to explore further the scaling behavior of $ \langle R_g^{2} \rangle_s $ with $N$. Note that, in absence of self-propulsion of the beads,  for the passive polymer this corresponds to an equilibrium state.  In this case, the spatial extension of the polymer is related to $N$ via the scaling form \cite{rubin_poly}
\begin{equation}\label{rg_scl}
		\langle R_g^2 \rangle_s \sim N^{2\nu}\,,
\end{equation}
where $\nu$ is known as the Flory exponent. For a passive chain with self-avoidance, the Flory approximation yields for extended coil conformations the value of $\nu=3/(d+2)$, with $d$ being the spatial dimensionality. For $d=3$, $\nu=3/5=0.6$ agrees well with the precise self-avoiding random walk exponent $\nu=0.58759700(40)$ \cite{clisby_16}. On the other hand, $\nu = 1/d$,  if the conformation of the chain is a globular one.  In Fig.~\ref{rgsq_ss_N} we plot $ \langle R_g^{2} \rangle_s$ versus $N$ for a few values of $Pe$. For the passive as well as for lower activity, for which the final conformation is a globule, the exponent is $\nu \approx 1/3$. 
 As expected from Fig.~\ref{length_ss_pe}(a), for the higher activities, the values of $\langle R_g^{2} \rangle_s$ are much larger compared to the other cases.  
 For both of them, data look consistent with a power-law behavior with the exponent  $\nu \approx 3/5$ similar to that for a self-avoiding random walk. This quantitatively  confirms that for higher activities the \textit{pseudo-equilibrium} steady-state conformations of the polymer are extended coils.

\par
\begin{figure}[t!]
	\centering
	\includegraphics*[width=0.45\textwidth]{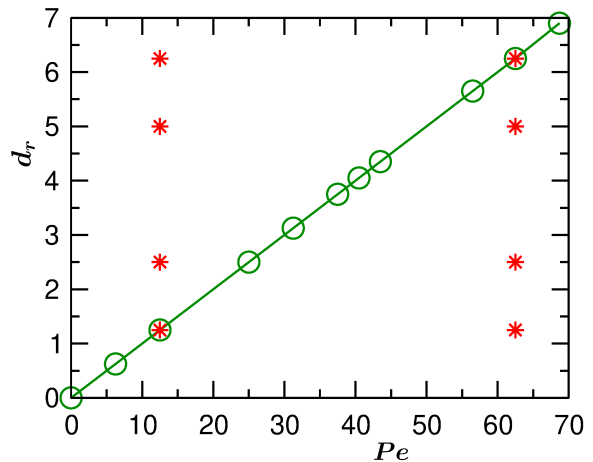}
	\caption{\label{schm_dr_pe} Plot showing the variation of $d_r$ versus $Pe$. The line has a slope $k_BT/\epsilon=0.1$. The stars mark the corresponding steady-state conformations of the polymer shown in Fig.~\ref{pe_dr_cons}.}
\end{figure}

Finally we investigated whether the basis for such a conformational change is embedded in our protocol used for changing the activity. In this regard, we  try to understand the relative importance of different energy scales present in the system. For the ABPo, along with the ``ballistic'' energy $f_p\sigma$ and the thermal energy $k_BT$, also the interaction energy $\epsilon$ is important. Alongside with $Pe=f_p\sigma/k_BT$, already defined as the activity parameter, one can define another dimensionless ratio  $d_r=f_p\sigma/\epsilon$.  Within the so far employed measurement protocol of changing $f_p$ at fixed $T$ and $\epsilon$ (keeping $\sigma$ fixed by default), not only $Pe$ but also $d_r$ changes.  In Fig.~\ref{schm_dr_pe} we plot the variation of $d_r$ versus $Pe$ while changing $f_p$ for fixed values of $\epsilon=1$ and $T=0.1$.  $d_r$ and $Pe$ are linearly related as
\begin{equation}\label{slope}
d_r =\frac{k_BT}{\epsilon}Pe\,,
\end{equation}
with a slope $k_BT/\epsilon=0.1$. The origin with both $d_r=Pe=0$ corresponds to the passive polymer case.

\par
\begin{figure}[t!]
	\centering
	\includegraphics[width=0.48\textwidth]{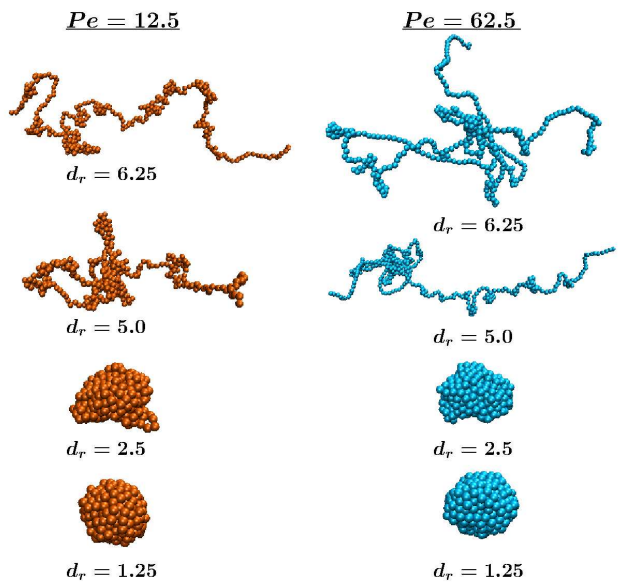}
	\caption{\label{pe_dr_cons} Steady-state conformations of the active polymer with a low ($Pe=12.5$) and a high activity ($Pe=62.5$) for four different values of $d_r$ for $N=380$.}
\end{figure}

 
In this context, we discuss another possible protocol: If one changes $Pe$ by varying the temperature $T$ for fixed values of $f_p$ and $\epsilon$, then $d_r$ remains unchanged. In this case one moves along a line parallel to the $Pe$ axis for any fixed $d_r$. Thus it is not possible to see the effect of the variation of $d_r$ on the polymer conformations. 
To investigate the effect of increasing $d_r$ in such a protocol, we show in Fig.~\ref{pe_dr_cons} the steady-state conformations for  different values of $d_r$ for a low and a high value of  $Pe$, i.e., $Pe=12.5$ and $62.5$. 
To perform these simulations we fix $f_p=2.0$.  Then the temperatures are fixed at $T=0.1$ and $0.02$ to set the $Pe$ values at $12.5$ and $62.5$, respectively. For each of them $d_r$ was changed by varying the interaction strength $\epsilon$. 
For both values of $Pe$, the conformations change from an  extended coil to a globule with decreasing $d_r$, i.e., increasing $\epsilon$, as observed from Fig.~\ref{pe_dr_cons}.  We also checked whether for the high $Pe$ case with small values of $d_r$ the globule conformation is generic or not for other combinations of the energy scales. For a higher $f_p=10.0$ with $T=0.1$ and $\epsilon=5.0$ (leading also to $Pe=62.5$ and $d_r=1.25$) we find that the steady state is also a globule. Now it becomes clearer that $Pe$ is not the only control parameter. Instead, irrespective of the choice of activity strength $f_p$, its relative importance compared to the interaction strength $\epsilon$, defined by $d_r$, helps in  determining the steady-state conformations. 
Since our way of changing $Pe$ also changes the value of $d_r$ we observe the globule to coil transition of ABPo with the variation of $Pe$. 

\section{ Conclusion}
In this paper, we have studied the kinetics of a flexible polymer consisting of active Brownian beads in a poor solvent. 
We considered a low enough temperature to minimize the effect of thermal noise on the dynamics of ABPo. The self-propelling force is varied to increase the activity while keeping all other parameters fixed.  One sees that with smaller activities, the conformations are  qualitatively similar to those for the passive case, and the final conformations are globules. 
The exponent for the scaling of relaxation time for small $Pe$ becomes lower indicating a faster dynamics for globule formation with increasing $Pe$. However,  with much higher activities the conformations become significantly different and the radius of gyration increases with time. This indicates that the polymer becomes more extended.
Variation of the gyration radius for the steady-state conformations versus the chain length shows that with increasing activity the corresponding Flory exponent changes from $1/3$ to $3/5$. This indicates a transition of the polymer from a globule to a self-avoiding random walk. 
\par 
We have understood such a conformational change on the basis of interplay among the three energy scales present in the system. According to our protocol, while changing the self-propelling force $f_p$, then along with the variation of the $Pe$ defined as the ratio of the active or ``ballistic'' energy $f_p\sigma$ and the thermal energy $k_BT$, another dimensionless ratio $d_r=f_p\sigma/\epsilon$, where $\epsilon$ is the interaction strength, also changes.  Our analyses confirm that the ratio $d_r$ plays an important role, driving a change of the steady-state conformation of the ABPo  from a globular to the coil state even though the condition is a poor solvent.  It will be worthwhile to look at the entire phase diagram for the steady-state conformations in the $Pe$-$d_r$ plane. This we intend to investigate in a future work.
\par 
Here we considered the Langevin equation in its overdamped limit to mimic the polymer moving in a viscous medium. For active particles one already observed significant differences in the clustering properties when replacing the overdamped Brownian dynamics with the underdamped Langevin equation \cite{lowen_20}. For the active polymer also it can be interesting to look at its properties while the  dynamics is mediated via the underdamped Langevin equation which features inertial effects.  In our model the ratio of the translational and rotational diffusion constants as well as the self-propulsion force on each bead are chosen as a constant throughout the simulation. 
But for real systems, in any typical biological environment, the active forces on the beads can have spatial and temporal dependencies  as all monomers may not experience the same force at all times. The dependence of particle motility on its local density has already been considered for the Vicsek model \cite{farrell_12}. By taking such a phenomenon into consideration for the ABPo also, this can provide more insights into its typical  conformations for real situations.  For the ABPo one can also take into account the effect of  hydrodynamics which arises due to interaction among the beads and the solvent particles \cite{winkler_20,gomez_20}. 

\section{Acknowledgment}
This project was funded by the Deutsche Forschungsgemeinschaft (DFG, German Research Foundation) 
under Grant No.\ 189\,853\,844--SFB/TRR 102 (Project B04). It was further supported by the Deutsch-Franz\"osische Hochschule (DFH-UFA) 
through the Doctoral College ``$\mathbb{L}^4$'' under Grant No.\ CDFA-02-07, the Leipzig Graduate School of Natural Sciences ``BuildMoNa'', 
and the EU COST programme EUTOPIA under Grant No.\ CA17139.

\end{document}